 \newcommand{\ketbra}[1]{\mathopen{|}#1\mathclose{\rangle}\hspace{-0.25em}\mathopen{\langle}#1\mathclose{|}}

\documentclass[aps,prl,twocolumn,floatfix,nopacs,longbibliography]{revtex4-1}
\usepackage{graphicx,xcolor,amsfonts,amssymb,amsmath,enumerate,upgreek,tabularx,multirow,bm,braket,comment}
\definecolor{link}{RGB}{46,46,145} % APS link color
\usepackage[colorlinks=true,allcolors=link]{hyperref}

\newcommand{\subfig}[2]{%
    {#1} \vtop{
  \vskip0pt
  \hbox{#2}
}}

\begin{document}

\title{Error-correction properties of an interacting topological insulator}

\author{Amit Jamadagni}
\email{amit.jamadagni@itp.uni-hannover.de}
\affiliation{Institut f\"ur Theoretische Physik, Leibniz Universit\"at Hannover, Appelstra{\ss}e 2, 30167 Hannover, Germany}
\author{Hendrik Weimer}
\affiliation{Institut f\"ur Theoretische Physik, Leibniz Universit\"at Hannover, Appelstra{\ss}e 2, 30167 Hannover, Germany}

\begin{abstract}

  We analyze the phase diagram of a topological insulator model
  including antiferromagnetic interactions in the form of an extended
  Su-Schrieffer Heeger model. To this end, we employ a recently
  introduced operational definition of topological order based on the
  ability of a system to perform topological error correction. We show
  that the necessary error correction statistics can be obtained
  efficiently using a Monte-Carlo sampling of a matrix product state
  representation of the ground state wave function. Specifically, we
  identify two distinct symmetry-protected topological phases
  corresponding to two different fully dimerized reference
  states. Finally, we extend the notion of error correction to
  classify thermodynamic phases to those exhibiting local order
  parameters, finding a topologically trivial antiferromagnetic phase
  for sufficiently strong interactions.

\end{abstract}
  
\maketitle

The classification of topological phases beyond the Landau symmetry
breaking paradigm remains an outstanding challenge in many-body
physics since the discovery of the topological origin of the integer
quantum Hall effect almost 40 years ago \cite{Thouless1982}. While the
non-interacting case is well understood in terms of topological
invariants \cite{Kohmoto1985}, giving rise to a plethora of
topological insulators and superconductors \cite{Kitaev2009},
counterexamples to succesful classification can be found in the case
of interacting systems \cite{Fidkowski2010}.

More recently, many-body topological invariants have been proposed to
classify one-dimensional phases with symmetry-protected topological
(SPT) order \cite{Haegeman2012,Pollmann2012,Elben2020}. However, the
possibility to acquire nonzero values even for topologically trivial
phases prevents a direct identification in terms of a topological
order parameter, meaning that succesful classification of phases
requires a complete set of invariants \cite{Chen2011}. To overcome
these challenges, we have recently introduced an operational
definition of topological order based on the ability of a system to
perform topological error correction \cite{Jamadagni2020}. While the
notion of topological error correction is chiefly motivated by quantum
memories such as the toric code exhibiting intrinsic topological order
\cite{Kitaev2003}, topological qubits based on Majorana fermions are
also prominently found within one-dimensional topological
superconductors \cite{Kitaev2001,Leijnse2012,DasSarma2015}, hinting at
a possible generalization of the operational definition.

In this Letter, we apply the operational definition in the context of
the Su-Schrieffer-Heeger (SSH) model \cite{Su1980}, a paradigmatic
model for a one-dimensional topological insulator \cite{Asboth2016},
whose bosonic variant has also recently been realized experimentally
using ultracold Rydberg atoms \cite{deLeseleuc2019}. While the SSH
model has so far not been discussed in the context of topological
error correction, we show that such a formulation can be readily found
by defining errors in terms of perturbations of the fully dimerized
limits of the model. We discover two distinct types of errors,
describing density and phase fluctuations, respectively, showing that
only the former are necessary to describe the topological phase
transition in the noninteracting case. We analyze the model
numerically in terms of a Monte-Carlo sampling of the error correction
procedure based on a matrix product state (MPS) calculation of the
ground state.  In particular, we find that both phases of the
noninteracting SSH model are topologically ordered, corresponding to
two different choices of the unit cell, which is also supported by the
appearance of distinct bulk topological invariants for the two
choices. Finally, we consider an extension of the bosonic SSH model
including antiferromagnetic interactions \cite{Elben2020}, where we
see that SPT order persists for finite interactions strengths before
an antiferromagnetic phase finally takes over.

\begin{figure*}[t]
\begin{center}
 \begin{tabular}{lp{0.25cm}c}
    \subfig{(a)}{\hspace{0.25cm}\includegraphics[width=.45\linewidth]{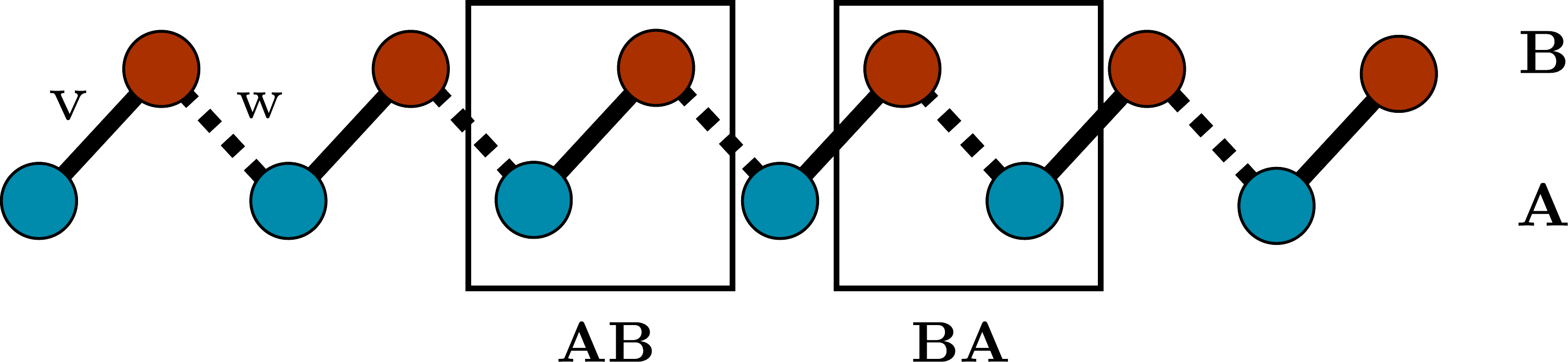}}\vspace{0.2cm}
    &&
    \multirow{2}{*}{\subfig{(c)}{\includegraphics[width=.45\linewidth]{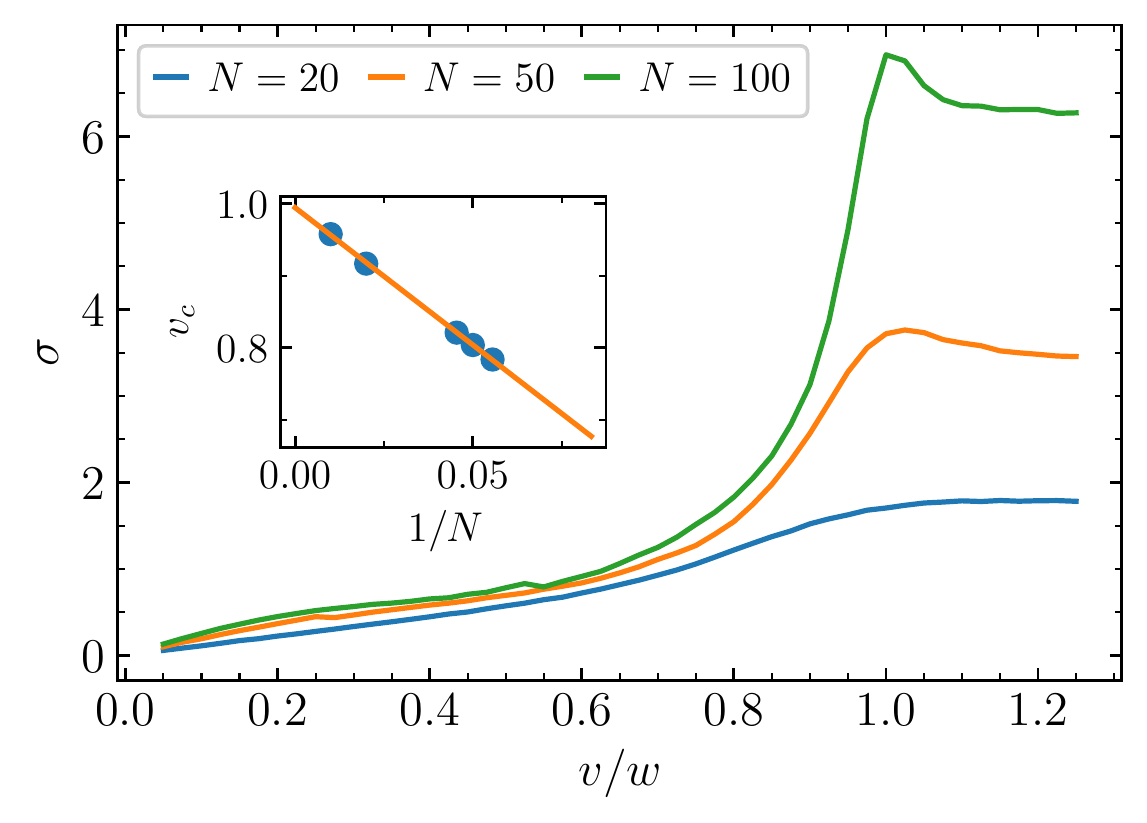}}}\\
    \subfig{(b)}{\hspace{0.25cm}\includegraphics[width=.45\linewidth]{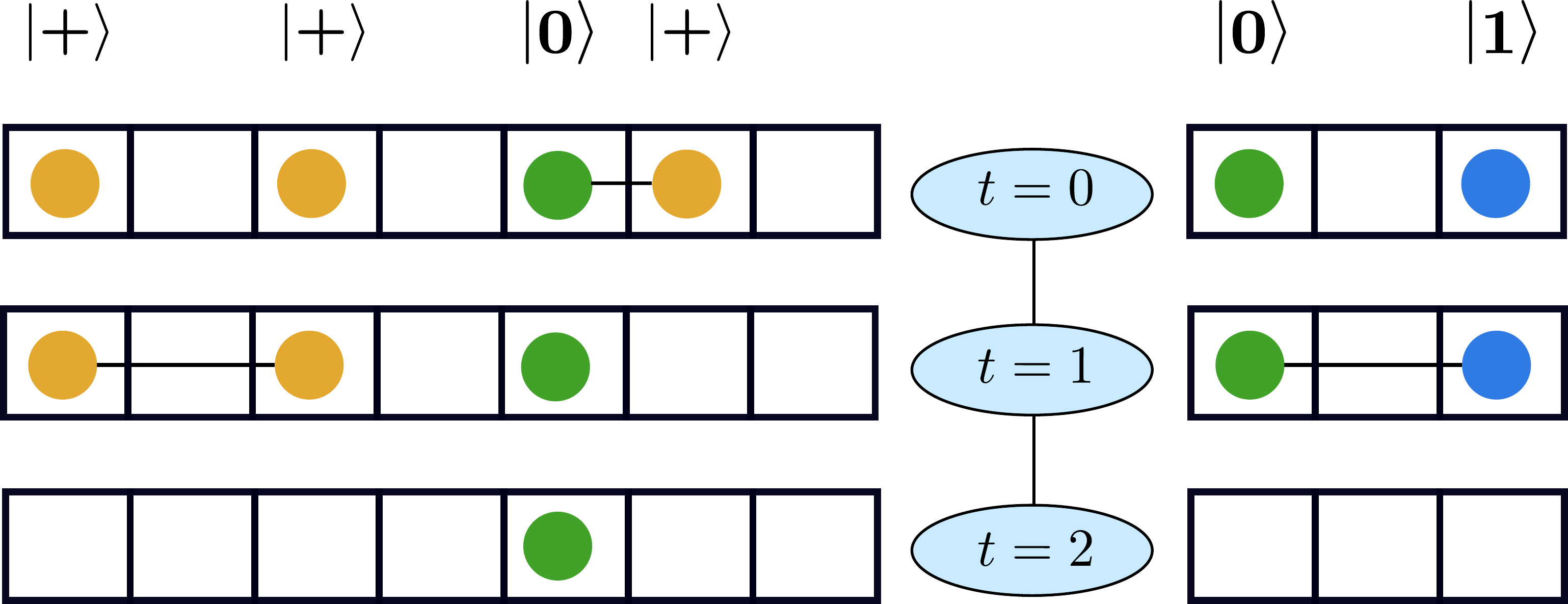}}
  \end{tabular}
  \vspace{0.75cm}
  \caption{(a) The SSH model can be understood as two alternating
    sites $A$ and $B$ coupled by bonds with interaction strength $v$
    and $w$, respectively. For periodic boundary conditions, there are
    two equivalent unit cells, denoted as $AB$ and $BA$. (b) Error
    correction for the SSH model. First (left), phase fluctuations
    (yellow) are corrected by either fusing them pairwise or absorbing
    them into a density fluctuation. In the second stage (right), the
    remaining density fluctuation are removed by fusing holes (green)
    and particles (blue). Fusion operations at the step $t$ are
    indicated by horizontal lines. The total circuit depth is the sum
    of the number of steps the two circuits require to return to the
    reference state. (c) Standard deviation $\sigma$ of the circuit
    depth $\sigma$ for the SSH model for the reference state $\ket{\psi_{AB}}$ 
    for different system sizes. Finite size scaling of peak of the susceptibility
    $\chi_v = \partial \sigma/\partial v$ (inset) yields a critical
    value of $v_c/w=1.00(1)$.}
\label{fig1}
\end{center}
\end{figure*}

\emph{SSH model.---} We consider the hardcore boson variant of the SSH model with the Hamiltonian being defined
on a 1D chain consisting of $N$ spin-1/2 particles as
\begin{align}\label{eq1}
H_0 = v\sum\limits_{i=1}^{N/2}\sigma_{-}^{2i-1}\sigma_{+}^{2i} 
+ w\sum\limits_{i=1}^{N/2-1}\sigma_{-}^{2i}\sigma_{+}^{2i+1} + \text{h.c.},
\end{align}
with the spin creation and annihilation operators satisfying the
commutation relation $[\sigma_-,\sigma_+] = \sigma_z$ in terms of the
Pauli spin matrix $\sigma_z$. In the case of open boundary conditions,
there is one $w$ link less than there are $v$ links, while for periodic
boundary conditions, the number of links are equal for both types. In
the latter case, the system is translationary invariant and can be
solved by mapping onto free fermions using a Jordan-Wigner
transformation and partitioning the system into unit cells of two
sites, which can be done in two different ways, see Fig.~1a. Since the
model is invariant under exchanging the unit cell and exchanging $v$
and $w$ at the same time, fixing the unit cell is similar to fixing a
gauge. For the $AB$ unit cell, performing a Fourier transform gives
rise to the band Hamiltonian $H(k) = d_{x}(k)\sigma_{x} +
d_{y}(k)\sigma_{y}$ with $d_{x}(k) = v + w\cos k$ and $d_{y}(k) =
w\sin k$, with the spin variable referring to the $A$ and $B$ sites
of a single unit cell \cite{Asboth2016}. Its eigenenergies are given
by $E(k) = |v + e^{-ik}w|$. From the energy spectrum we note that for
$v<w$ and $v>w$, the band gap is finite resulting in insulating phases
while at $v=w$ we have a conductor due to the closing of the band
gap. We can see that this closing of the gap is due to the presence of
a phase transition between distinct topological phases by considering
two different topological invariants corresponding to the choice of
the unit cell. Specifically, we consider the winding number
\begin{align}
\nu = \frac{1}{2\pi i}\int_{-\pi}^{\pi}dk \frac{d}{dk}\log h(k)
\end{align}
where $h(k) = d_{x}(k) - id_{y}(k)$ \cite{Asboth2016}. Choosing the
$AB$ as the unit cell, we have $\nu_{AB}=1$ for $v<w$ and $\nu_{AB}=0$
for $v>w$. Due to the presence of the $v \leftrightarrow w$ symmetry, we
immediately see that we have $\nu_{BA} = 0$ in the former and
$\nu_{BA} = 1$ in the latter case.

\begin{figure*}[t]
\begin{center}
  \begin{tabular}{cp{0.5cm}c}
    \subfig{(a)}{\includegraphics[width=.4\linewidth]{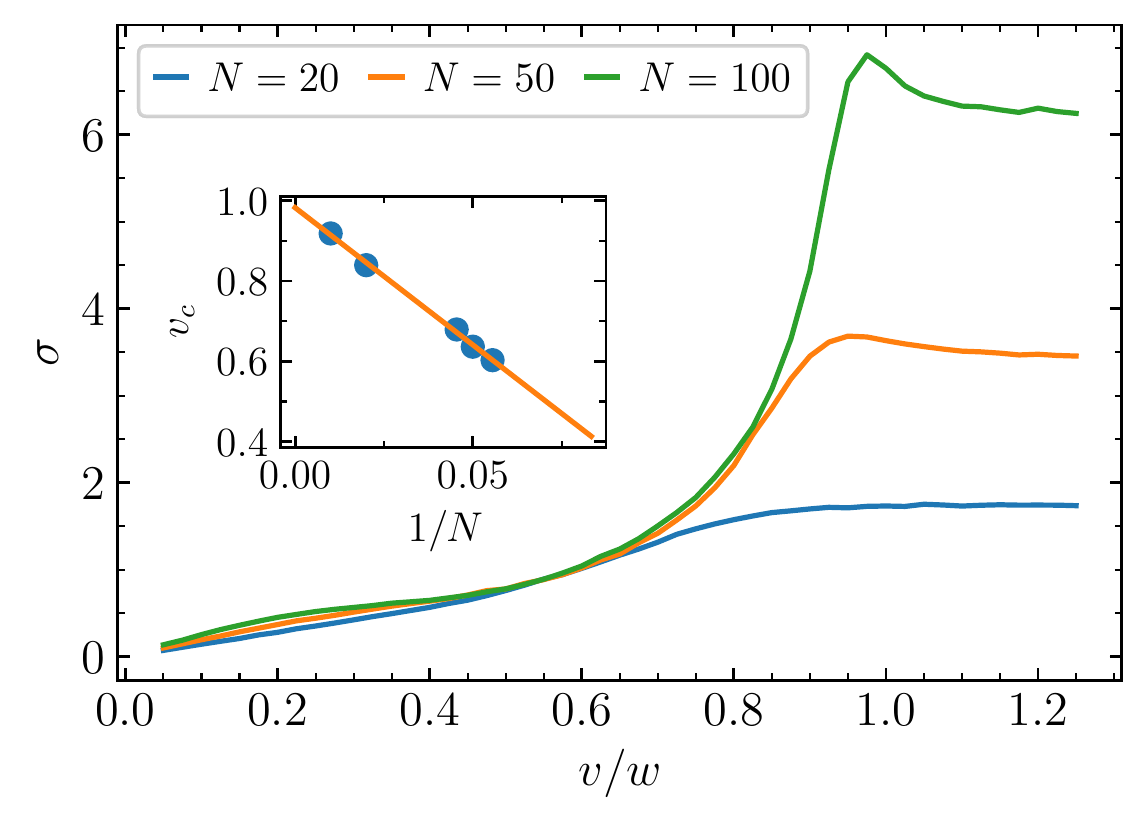}}
    &&
    \subfig{(b)}{\includegraphics[width=.4\linewidth]{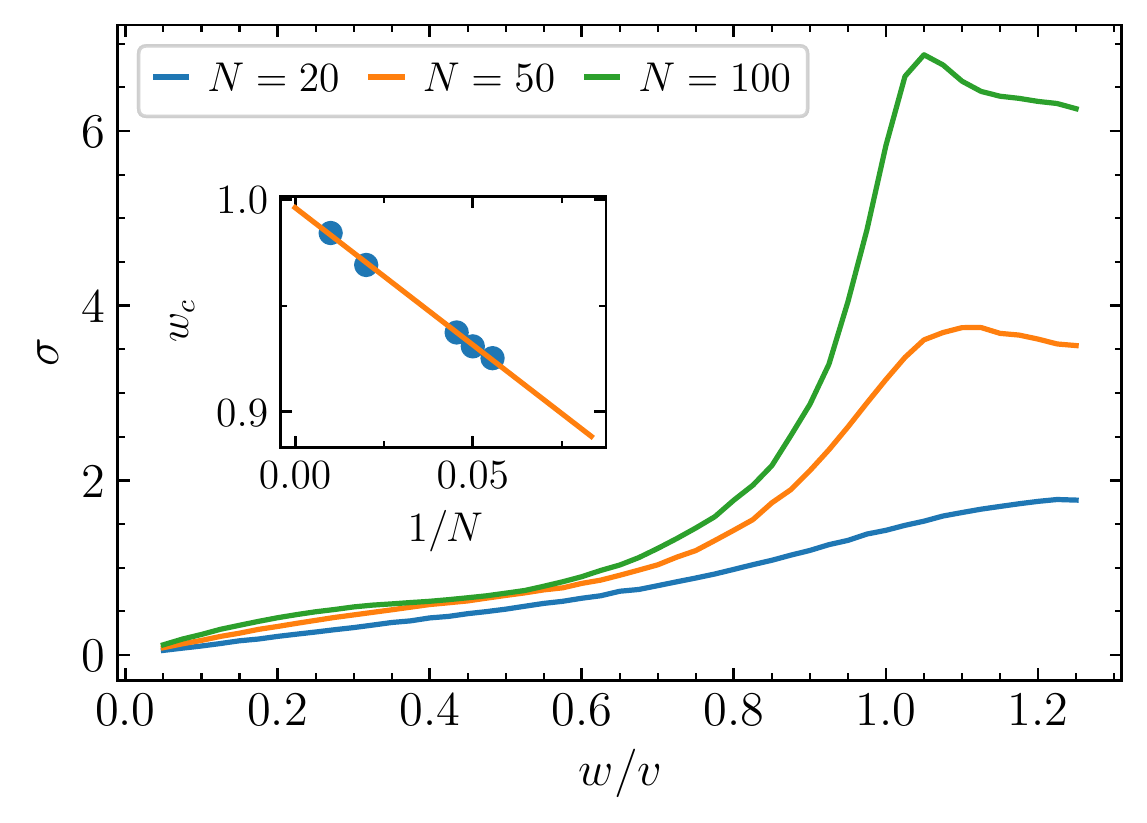}}
  \end{tabular}
\end{center}
\caption{Standard deviation $\sigma$ of the circuit depth $\sigma$ for the reference state 
$\ket{\psi_{AB}}$ (a) and $\ket{\psi_{BA}}$ (b). The insets show the finite size scaling 
resulting in $v_c/w=0.98(1)$ and $w_c=1.00(1)$, respectively.}
\label{fig2}
\end{figure*}

Interestingly, the error basis can be also used to provide additional
insight into the topological phase transition. Within perturbation
theory in $v/w$ (for the $AB$ unit cell), we find that phase
fluctuations correspond to higher oder processes compared to density
fluctuations. This means that we can neglect the $\ket{\bm{+}}$ state
in an effective low-energy description of the SSH model, arriving at a
spin-1 representation according to
\begin{equation}
  H_\text{eff} = \sum\limits_i w {S_i^z}^2 + v(S_i^+S_{i+1}^- + \text{H.c.}).
  \label{eq:heff}
\end{equation}
Here, $w$ takes the role of an uniaxial anisotropy, while $v$
describes a hopping of the spin excitations. Constructing such
effective models for the dynamics of errors is reminiscent of
effective Ising models describing the topological phase transition in
perturbed toric code models \cite{Trebst2007,Tagliacozzo2011}. The
phase diagram of this spin-1 model is well-known
\cite{Golinelli1992,Chen2003,Gu2009}, exhibiting a phase transition
between a large-$w$ phase corresponding to the fully dimerized limit
and a Haldane insulator at $v/w=1$. Note that the phase for $v>w$ is
also topologically ordered; this is another manifestation of the $v
\leftrightarrow w$ symmetry relating the two phases to each other.

\emph{Error correction properties.---} Although the non-interacting
SSH model is exactly solvable, it is instructive to numerically study
its error correction properties, which will serve as a base to
investigate the interacting case. For this, we turn to the operational
definition of topological order \cite{Jamadagni2020}, which relates
the existence of a phase transition to the divergence of the depth of
an appropriate error correction circuit with respect to a particular
reference state. For the SSH model, the reference states can be readily
identified as the ground states in the fully dimerized limit given by
$v=0$ or $w=0$, respectively, i.e.,
\begin{equation}
  \ket{\psi}_{AB/BA} = \frac{1}{\sqrt{2}}\prod\limits_{i\in B/A} (\ket{0}_i\ket{1}_{i+1}-\ket{1}_i\ket{0}_{i+1}).
\end{equation}
The errors with respect to these reference states can then be found by
considering a complete basis of possible excitations. For a single
bond between two unit cells, we can denote the error-free state as
$\ket{\bm{-}} = (\ket{01}-\ket{10})/\sqrt{2}$. Additionally, we can
identify particle and hole excitations indicated by the states
$\ket{\bm{1}} = \ket{11}$ and $\ket{\bm{0}} = \ket{00}$, respectively,
as well as phase fluctuations given by $\ket{\bm{+}} =
(\ket{01}+\ket{10})/\sqrt{2}$. Repeating this procedure over the
entire system will then define a unitary transformation allowing us to
express any state in this error basis.

Having specified the reference state and a complete set of errors, we
now turn to the actual error correction procedure. In the following,
we assume that the system has been measured in the error basis,
yielding a classical string of errors. As the topological phases of
the SSH model are SPT phases \cite{Micallo2020}, we need to ensure
that the error correction circuit cannot perform operations that
violate these symmetries \cite{Jamadagni2020}. For example, performing
an operation that corrects a single $\ket{\bm{+}}$ error directly to
the $\ket{\bm{-}}$ state would be a violation of the chiral
symmetry. A further minor complication arises due to the fact that
phase errors $\ket{\bm{+}}$ can arise as higher-order processes from
density errors. This requires to correct phase errors before density
errors, as otherwise correcting the density errors first can lead to
dangling phase errors, which manifest themselves as spuriously
diverging circuit depths. Taking these considerations into account, we
arrive at the error correction procedure depicted in Fig.~1b: (i) We
assign a walker to each measured error that searches its surrounding
sites for the presence of other errors, switching between left and
right with increasing distance from the initial position
\cite{Jamadagni2020}. (ii) Phase errors get corrected by either fusing
them pairwise or with a particle-hole error. (iii) Finally, density
errors are corrected by fusing particles and holes. The depth of the
circuit is then given by the total number of steps required to correct
the system to the reference state.

\emph{Monte-Carlo sampling of matrix product states.---} As the
circuit depth corresponds to a highly nontrivial string operator, it
is prohibitive to compute its expectation value from the exact
solution of the model. Therefore, we turn to MPS calculations of the
ground state using the ITensor library \cite{Fishman2020} up to
$N=100$ sites. However, even within a matrix product state formalism,
efficient computation of arbitrary string operator expectation values
is in general impossible, we turn to a Monte-Carlo sampling of the
error measurements instead \cite{Han2018}. For this, we start with a
MPS $\ket{\psi}$ by calculating the probabilities to measure any of
the basis states of the error basis $\ket{\alpha}_{1,2}$ of the first
unit cell in terms of the expectation value of the associated
projection operators $P_\alpha = \ketbra{\alpha}$. Subsequently, we
draw a uniformly distributed random number to select the measurement
result according to the probabilities $\langle
P_\alpha\rangle$. Denoting the measurement result by $\alpha_{1,2}$,
we can update the MPS according to $\ket{\psi'} =
\mathcal{N}P_{\alpha_{1,2}}\ket{\psi}$, with $\mathcal{N}$ referring
normalization of the MPS, yielding the MPS conditional on the
measurement result. Continuing the procedure over the entire system
results in a string $\alpha_{1,2}\alpha_{2,3}\ldots$ of the entire
error configuration. Sampling over a large number of measurement
outcomes and calculating the circuit depth for each outcome will then
lead to an accurate estimation of the mean circuit depth or higher
order moments such as the standard deviation.

Figure 1c shows the behavior of the circuit depth for the
non-interacting SSH model. Here, we focus on the standard deviation
$\sigma$ of the depth as it exhibits slightly better finite scaling
results. Due to the $v \leftrightarrow w$ symmetry of the model, it is
sufficient to study only. We clearly see a
divergence of the circuit depth around the critical value of $v/w=1$,
signaling the phase transition. Furthermore, finite size scaling
reveals the critical point as $v_c/w = 1.00(1)$, i.e., the error
correction procedure reproduces the quantitatively correct result.

\begin{figure*}[t]
\begin{center}
  \begin{tabular}{cp{0.01mm}cp{0.01mm}c}
    \subfig{(a)}{\includegraphics[width=.29\linewidth]{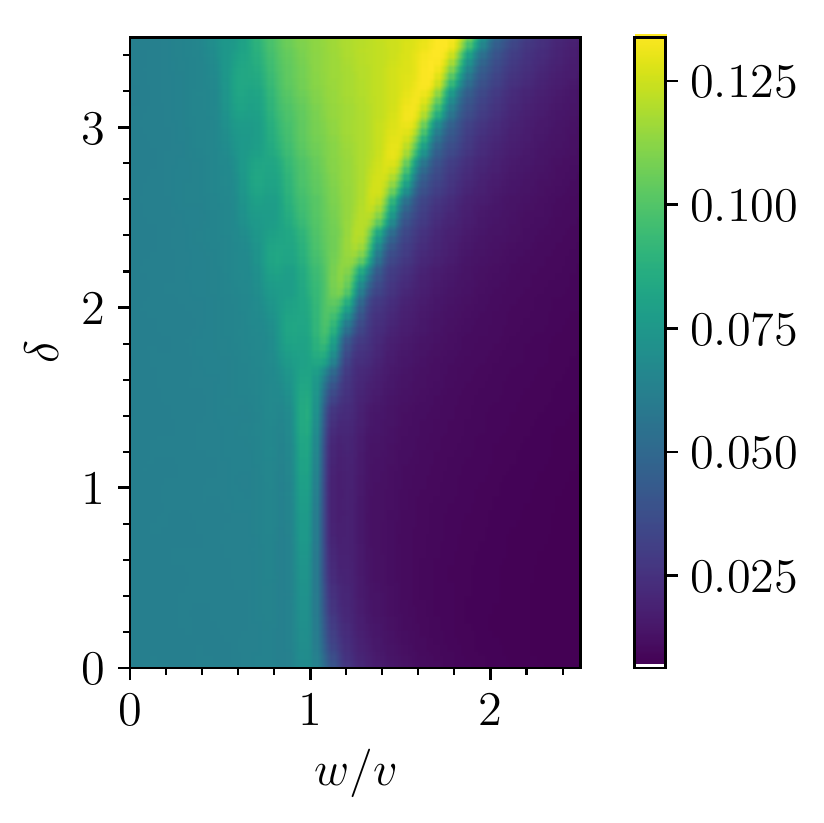}}
    &&
    \subfig{(b)}{\includegraphics[width=.29\linewidth]{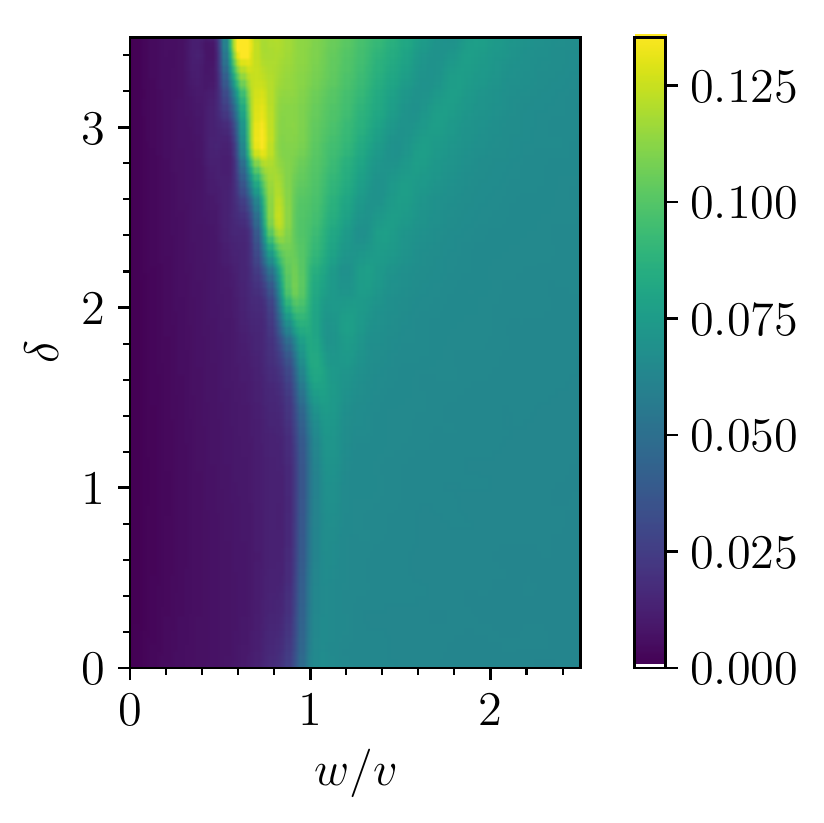}}
    &&
    \subfig{(c)}{\includegraphics[width=.29\linewidth]{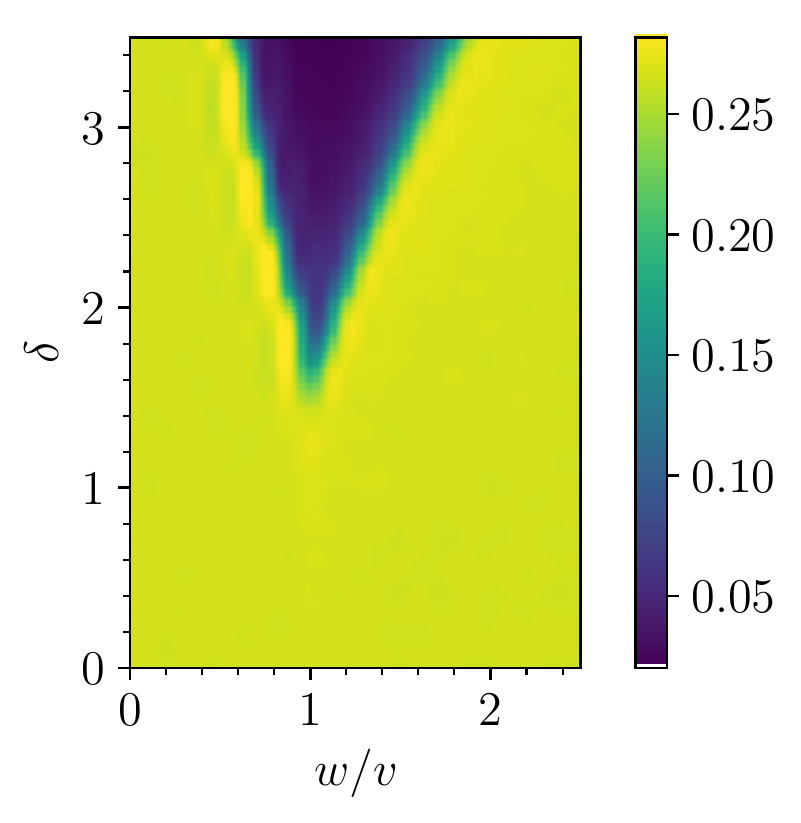}}\\
%    &&
%    \subfig{(d)}{\includegraphics[width=.3\linewidth]{Figs/afm_ec.pdf}}
  \end{tabular}
\end{center}
\caption{Phases of the extended SSH model, calculated by the circuit
  depth to correct the ground state to the two SPT ordered reference
  states $\ket{\psi_{AB}}$ (a) and $\ket{\psi_{BA}}$ (b), as well as
  to the antiferromagnetic state $\ket{\psi_{AF}}$ (c) for $N=100$
  sites. The dark areas indicate a vanishing of the normalized circuit
  depth $\sigma/N$, showing that the ground state is in the same phase
  as the reference state. All areas of short circuit depths are
  mutually exclusive and span the entire parameter range.}
  \label{fig3}
\end{figure*}

\emph{Open boundary conditions.---} So far, we have discussed the SSH
model in the context of periodic boundary conditions. However, most
interest in the SSH model lies in the realization of open boundary
conditions due to the appearance of robust edge modes capable to store
quantum information \cite{Asboth2016}. Implementation of the necessary
error correction circuits can be done in a straightforward way from
the periodic case. However, the possible degeneracy implies that the
parity of the ground state is not well-defined. Therefore, we ignore
the state of the two edge spins in the case of $\ket{\psi_{AB}}$
as the reference state.

Figure 2 shows the MPS simulation results for both reference states
$\ket{\psi_{AB}}$ and $\ket{\psi_{BA}}$. As in the case of periodic
boundary conditions, we observe a phase transition between the two
phases at $v/w=1$. However, since the introduction of open boundaries
breaks the $v \leftrightarrow w$ symmetry, the identification of both
phases as SPT phases deserves further discussion. For the $v<w$, this
identification is straightforward, as the edge mode can be used to
encode a topological qubit, whose logical state is preserved under the
error correction circuit. However, this argument does not hold for
$v>w$ as the ground state is unique. Nevertheless, we can establish
the phase being SPT ordered by inspecting the reference state
$\ket{\psi_{BA}}$. Since the reference state is a producFigs/t state of all unit
cells, it is sufficient to look at a single unit cell. A state is SPT
ordered, if there is no set of symmetry-preserving local unitaries
that transform the state into a product state \cite{Chen2010}. To
identify the possible unitaries on a single unit cells, we note that
all accessible states have to be in the same symmetry sector as the
$\ket{\bm{-}}$ state with respect to the chiral symmetry and the
$U(1)$ symmetry corresponding to particle number conservation
\cite{Micallo2020}. Crucially, there is no other state that fulfills
these criteria. This means that there is no symmetry-preserving
circuit that can transform the state $\ket{\psi_{BA}}$ to a product
state and hence this phase must be SPT ordered.

Additionally, it is instructive to look at the effective low energy
Hamiltonian (\ref{eq:heff}) again to obtain insight into the breaking
of the ground state degeneracy by the topological phase transition
\cite{Jamadagni2018}. For this, we consider the effective Hamiltonian
on the first site of the lattice, which in a mean-field decoupling is
given by $H_1 = vS_1^-\langle S_2^+\rangle + \text{H.c.}$. Within a
mean-field decoupling, $\langle S_2^+\rangle$ is nonzero only for
$v>w$, opening a gap between the edge modes above the
transition. While this simple mean-field decoupling is unable to
correctly describe the Haldane insulator, one can expect that this
argument also holds within a more refined treatment
\cite{Kennedy1992}.

Figs/
\emph{Extended SSH model.---} Let us now go beyond the noninteracting
case and study an extension of the SSH model including
antiferromagnetic interactions \cite{Elben2020}, given by the Hamiltonian
\begin{equation}
  H = H_0 + \frac{\delta}{2}\left[v\sum\limits_{i=1}^{N/2}\sigma_{z}^{2i-1}\sigma_{z}^{2i} 
    + w\sum\limits_{i=1}^{N/2-1}\sigma_{z}^{2i}\sigma_{z}^{2i+1}\right].
\end{equation}
While we discuss the bosonic version of the model here, we would like
to note that one can also study an equivalent fermionic version
including a chemical potential and a nearest-neighbor interaction,
where the chemical potential is tuned such that the particle-hole
symmetry of the SSH model is preserved.

In the limit of large $\delta$, it is evident that the terms in $H_0$
are irrelevant and the ground state is an Ising antiferromagnet. It is
easy to construct a reference state for this phase, as it is simply a
classical state $\ket{\psi_{AF}}=\ket{010101\ldots}$. Errors are given by domain wall
excitations located on the bonds between two sites occurring when the
spin state of these sites is identical. Local spin flips always create
these excitations in pairs, hence the error correction is given by the
pairwise fusion of all domain walls. Note that in contrast to the
previous reference state, we have only one elementary excitation
instead of two. This can be attributed to the fact that an Ising
antiferromagnet can only reliably store \emph{classical} information,
while the edge mode in the SSH model can store \emph{quantum}
information, i.e., while an Ising antiferromagnet can correct bit-flip
errors, it does not correct phase errors.

In Figure 3, we show the error correction properties of the extended
SSH model as a function of $w/v$ and $\delta$ for all three reference
states $\ket{\psi_{BA,AB,AF}}$. We can clearly see that the areas of
short circuit depths are mutually exclusive and span the entire
parameter range, i.e., the error correction approach can be
successfully employed to determine the complete phase
diagram. Additionally, the phase boundaries are in excellent
quantitative agreement with those obtained using an approach based on
the application of random quantum gates \cite{Elben2020}.

In summary, we have explored the phase diagram of an interacting
topological insulator model based on the error correction properties
of the ground state. We show that this approach can successfully map
out the entire phase diagram, including the transition to an
antiferromagnetic phase exhibiting spontaneous symmetry
breaking. Finally, the operational character based on measurable
observables enables to directly detect topological order in future
experimental studies.

\begin{acknowledgements}
  We thank S.~Diehl, P.~Recher, and B.~Vermersch for fruitful
  discussions. This work was funded by the Volkswagen Foundation, by
  the Deutsche Forschungsgemeinschaft (DFG, German Research
  Foundation) within SFB 1227 (DQ-mat, project A04), SPP 1929 (GiRyd),
  and under Germany’s Excellence Strategy -- EXC-2123 QuantumFrontiers
  -- 390837967.
\end{acknowledgements}

\bibliographystyle{myaps}
\bibliography{sshp.bbl}

\end{document}